\documentclass[pra,bibnotes,twocolumn]{revtex4}
\usepackage{graphicx}
\begin{document}
\draft
\def\ds{\displaystyle}
\title{ Spontaneously broken particle-hole symmetry in photonic graphene with gain and loss }
\author{Z. Oztas, C. Yuce}
\address{Department of Physics, Eskisehir Technical University, Turkey }
\email{cyuce@anadolu.edu.tr}
\date{\today}
\begin{abstract}
We consider particle-hole symmetric photonic graphene with balanced gain and loss. We show that edge states with purely imaginary eigenvalues appear along the zigzag edge. We propose an idea that these edge states are protected by spontaneously broken particle-hole symmetry.  We discuss that the edge states are topological in the sense that the exceptional rings are robust against symmetry protecting disorder. If the disorder is too strong to restore the spontaneously broken particle-hole symmetry, then such protection of the edge states is lost. 
\end{abstract}
\maketitle

 \section{Introduction}
 
Graphene, which is a two dimensional  honeycomb lattice of carbon atoms, has attracted great attention in the last decade. Dirac cones appear in its band structure and hence conducting electrons in graphene move as if they are massless relativistic fermionic particles. Graphene is not a topological insulator since it is a two dimensional gapless system and its Chern number is zero. However, it has protected edge states in the sense that each of the Dirac cones has a Berry phase of $\pi$ and  $-\pi$, respectively. Therefore, the edge states in graphene are robust against weak symmetry protected perturbations \cite{grphnedge02}. Indeed, it was shown that the edge states survive until two Dirac cones with opposite Berry phase merge in such a way that their Berry phases cancel each other. We emphasize that this topological protection and the one in topological insulators that forbids backscattering should not be confused. There are three types of edges of graphene: the zig-zag, bearded, and armchair edges. Edge states were theoretically predicted \cite{predic1,grphnedge00,grphnedge01} and experimentally realized along zigzag edges in graphene \cite{predic2}. The idea of edge states in graphene structure is extended to photonics \cite{grphnedge03}. The artificial photonic graphene has some advantages to atomic graphene such as the absence of interaction between photons and easy manipulation of the lattice geometry. \\
Edge states have raised interest in various field of physics after the discovery of topological insulators. For instance, there is a growing interest in studies of edge states in non-Hermitian systems. Photonic systems with gain and loss are thought to present new physics inaccessible in the context of condensed matter. The non-Hermitian extensions of topological insulators and superconductors are emergent field of study. In this new subfield, researchers generally focus $\mathcal{PT}$ symmetric systems, where $\mathcal{P}$ and $\mathcal{T}$ are parity and time reversal operators, respectively \cite{borc1}. However, few paper have appeared in the literature on this topic in the first decade after the discovery of topological insulators. The main reason for this is that topological phase was believed not to be compatible with non-Hermitian systems \cite{PTop3,ekl56,PTop4,PTop1}. There is still no general theory available to explain topological phase in non-Hermitian systems. In 2015, it was theoretically predicted that stable topological phase exists in a non-Hermitian Aubry-Andre model  \cite{cem0001}. Two years ago,  stable topological zero energy edge states were observed in an experiment on a photonic lattice which consists of waveguides with staggered hopping amplitudes \cite{sondeney1}. In the last two years, many other papers have been published in the literature. Most of the papers deal with one dimensional problems \cite{sbt1,sbt3,sbt3ekleme,sbt3ekleme2,sbt3ekleme36,sbt3ekleme3,flotop11,sbt6,sbt11,sbt8,sbt2,sbt500}. For instance, complex extensions of the Su-Schrieffer-Heeger (SSH) model was shown to support topological zero energy states \cite{sbt4,cmyc}. Topological superconductors with gain and loss have also been investigated and Majarona modes in such systems have been explored \cite{sbt012a,sbt12,sbt16,sbt17,sbt7,sbt18}. Other interesting studies are Floquet topological insulators with gain and loss that appear in time-periodic systems \cite{sbt10,sbt15} and topological laser that are robust against any fabrication defects and local defects  \cite{laser01,laser02,laser03,laser04,laser06}. \\
It was shown that electron-hole pairs in graphene leads to particle-hole symmetry \cite{oneri1,oneri2,oneri3,oneri4}. In atomic graphene, electron-hole attraction poses problems from the topological insulating point of view. In this paper, we consider a photonic analogue of graphene \cite{photgrap,borc2}. By adding gain and loss to each sublattice, a non-Hermitian graphene structure can be obtained \cite{photgrap}. We consider this structure and show that edge states with purely imaginary eigenvalues appear along the zigzag edge of the system. We discuss that edge states don't have any symmetries while the bulk states are particle-hole symmetric. Some questions arise. Which symmetry protects these edge states? Do the eigenvalues of these edge states still resist to particle-hole symmetric perturbations? Here, we propose that spontaneously broken particle-hole symmetry protects the edge states along the zig-zag edge. This idea is new and such a protection is unique to non-Hermitian systems.

\section{Complex zig-zag edge states in graphene with balanced gain and loss}

The graphene structure is a honeycomb lattice in two spatial dimensions as shown in the Fig.1, where the red and blue sites denote the bipartite sublattices. The blue sites have only first neighbors to red sites and vice versa. Suppose now that we introduce balanced gain and loss to the blue and red sites, respectively. In this case, we obtain a non-Hermitian extension of graphene structure. Note that gain and loss are introduced as on-site imaginary potentials. Therefore, the corresponding non-Hermitian Hamiltonian has the form 
\begin{equation}\label{mythfncz2}
\mathcal{H} (k)=d_x(\textbf{k})~\sigma_x+d_y(\textbf{k})~\sigma_y+i~{{\gamma}}~ \sigma_z
\end{equation}
where $\ds{\sigma_x}$, $\ds{\sigma_y}$ and $\ds{\sigma_z}$ are Pauli matrices, $\gamma$ is the non-Hermitian degree and $\textbf{k}$ is the crystal momentum defined in the first Brillouin zone in $2D$ and $d_x$ and $d_y$ are given by \cite{predic1}
\begin{eqnarray}\label{mqasdfv2D}
d_x+id_y=e^{-ik_x }+2~e^{ i  \frac{3}{2} k_x}  ~ \cos(\frac{\sqrt{3}}{2}  k_y )  
\end{eqnarray}
This Hamiltonian is reduced to the well known standard Hamiltonian for graphene when $\ds{\gamma=0}$. Let us first discuss the symmetry properties for this Hamiltonian. One can see that the Hermitian graphene $\ds{({\gamma}=0)}$ is $\ds{\mathcal{PT}}$ symmetric
\begin{equation}\label{yudj2}
\sigma_x~ \mathcal{H}^{\star}(k_x,k_y) ~\sigma_x= \mathcal{H}(k_x,k_y)   
\end{equation}
where $\ds{\mathcal{PT} =  \sigma_x~  \mathcal{K} }$ with $\ds{( \mathcal{PT})^2 = 1}$ and $\ds{{\mathcal{K}}}$ is the complex conjugation operator. Note that both the parity and time reversal operators change the sign of $k_x$ and $k_y$. It is interesting to see that this $\ds{\mathcal{PT}}$ symmetry remains intact even in the presence of gain and loss, $\ds{\gamma\neq0}$. In addition to the parity-time symmetry, both the Hermitian and the non-Hermitian Hamiltonians have also particle-hole symmetry
\begin{equation}\label{yundunmu}
\sigma_z~ \mathcal{H}^{\star}(-k_x,-k_y) ~ \sigma_z= -\mathcal{H}(k_x,k_y)   
\end{equation}
where particle-hole symmetry operator reads $\ds{\mathcal{C}=\sigma_z~  \mathcal{K}  }$. The particle-hole symmetry implies that energy eigenvalues always appear in a pair for each $\textbf{k}$. An important difference between the Hermitian and non-Hermitian graphene structure is as follows: The $\ds{\mathcal{PT}}$ and $\ds{\mathcal{C}}$ symmetries always present for all $\textbf{k}$-values in the Hermitian system. However, they are spontaneously broken for some particular values of $\textbf{k}$ in the non-Hermitian system. If $\gamma$ is too large, the spontaneously broken region is covered in the whole Brillouin zone. Note that the spontaneous breaking of the $\ds{\mathcal{PT}}$ and $\ds{\mathcal{C}}$ symmetries occur simultaneously in the same domains. The spontaneous breaking of the $\ds{\mathcal{PT}}$ symmetry means that the eigenstates of the Hamiltonian are no longer the eigenstates of the $\ds{\mathcal{PT}}$ operator. This occurs when the energy eigenvalues become complex valued. The spontaneous breaking of the $\ds{\mathcal{C}}$ symmetry means that the eigenstates of the Hamiltonian are no longer switched under the $\ds{\mathcal{C}}$ operator. The particle-hole symmetry implies that 
\begin{equation}\label{yurgnmu}
\sigma_z~ \Psi_{\mp}^{\star}(-k_x,-k_y) =- \Psi_{\pm}(k_x,k_y) 
\end{equation}
where $\ds{\Psi_{\mp}(k_x,k_y)}$ are upper and lower eigenstates of the Hamiltonian (\ref{mythfncz2}). This relation is true as long as the particle-hole symmetry is not spontaneously broken, i.e., as long as the corresponding energy eigenvalues for $\ds{\Psi_{\mp}(k_x,k_y)}$ remain real valued. \\
\begin{figure}[t]
\includegraphics[width=9cm]{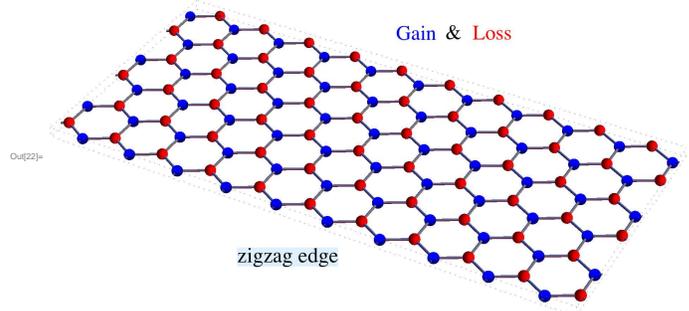}
\caption{ The graphene structure with two sublatttices, depicted as blue and red colors. Gain and loss are introduced into the blue and red sites, respectively. One can also see the zigzag edge of the graphene structure. It is periodic along the long direction and terminated along the short direction. }
\end{figure}
Spontaneously broken symmetries lead to appearance of exceptional points. To study exceptional points, let us write down the energy eigenvalues of the system. The upper and lower energy eigenvalues of (\ref{mythfncz2}) are given by $\ds{E_{\mp}=\mp\sqrt{d_x^2+d_y^2-{\gamma}^2 }}$. It is well known that the Hermitian graphene structure exhibits six Dirac points arranged in a regular hexagon. These points occur at six different $\ds{\textbf{k}^{\prime}}$ where the band gap closes: $\ds{d_x(\textbf{k}^{\prime})=d_y(\textbf{k}^{\prime})=0}$. In the non-Hermitian system, these Dirac points at $\ds{\textbf{k}^{\prime}}$ are no longer Dirac points since the $\ds{\mathcal{PT}}$ and $\ds{\mathcal{C}}$ symmetries are spontaneously broken at these points and the corresponding energy eigenvalues become purely imaginary. Note that these two symmetries are not spontaneously broken in the whole Brillouin zone, but only in the neighborhood of $\ds{\textbf{k}^{\prime}}$. At the border of this region around $\ds{\textbf{k}^{\prime}}$,  exceptional points occur, where the upper and lower eigenstates coalesce and the corresponding energy eigenvalues are equal to zero. The set of exceptional points forms an exceptional ring, which is governed by the equation $\ds{d_x(\textbf{k})^2+d_y(\textbf{k})^2=\gamma^2}$. The exceptional rings are not exactly circular and their centers are arranged in a regular hexagon. They are well separated for small values of $\gamma$ but start to get fused in pairs when $\gamma=1$. The merged exceptional ring spreads in the Brillouin zone as $\gamma$ is increased from $\gamma=1$ to $\gamma=3$. No exceptional ring exists and all the energy eigenvalues are complex valued for all $\textbf{k}$ if $\gamma > 3$.\\
The Hermitian graphene is known to support zero-energy edge states along the zigzag edge. The edge states arise since the Berry phase for a closed loop around one Dirac cone of the graphene lattice yields either $\pi$ or $-\pi$. But these edge states are not strictly topological since the total Berry's phase is zero. The edge states have zero energy and decay exponentially into the bulk of the lattice. Let us discuss what happens if the gain and loss are introduced into the system. Exceptional rings arise from the Dirac points if $\gamma$ is switched on and grow with increasing $\ds{\gamma}$. So, there are three distinctive regions depending on $\ds{\gamma}$: i-) small values of $\ds{\gamma}$, i.e., $\ds{\gamma<<1}$: The $\ds{\mathcal{PT}}$ and $\ds{\mathcal{C}}$ symmetries are spontaneously broken in small regions around each Dirac points. Therefore we expect that only the edge states have imaginary energy eigenvalues. ii-) intermediate values of $\ds{\gamma}$, i.e., $\ds{\gamma\sim1}$: We expect that the edge states and some of the bulk states have complex energy eigenvalues. iii-) large values of $\ds{\gamma}$, i.e., $\ds{\gamma>>1}$: the $\ds{\mathcal{PT}}$ and $\ds{\mathcal{C}}$ symmetries are spontaneously broken in the whole Brillouin zone for large values of $\ds{\gamma}$ and hence the edge states and all of the bulk states have complex energy eigenvalues. To confirm our discussion, we perform numerical calculations. For that purpose, the lattice is assumed to be periodic in the one direction but finite in the perpendicular direction. For the armchair edge termination one does not find any zero-energy edge states, hence we consider only zig-zag edge termination. The Fig-2 plots the real and imaginary parts of the energy eigenvalues for a small and intermediate values of the non-Hermitian degree: $\ds{\gamma=0.1}$ and $\ds{\gamma=1}$. As can be seen, there are only two states with complex energy eigenvalues for $\ds{\gamma=0.1}$. These two states are the edge states that exist on the zigzag edges between $k=\mp\pi$ and $k=\mp2\pi/3$. In this case, all bulk states have real valued energy eigenvalues. This can be used as a laser working along edges. On the other hand, not only edge states but also a couple of bulk states have complex energy eigenvalues for $\ds{\gamma=1}$ as can be seen from the  Fig-2. We perform one last numerical calculation for $\gamma=3$ and we find that all of the eigenstates states have purely imaginary energy eigenvalues. So far, we have studied the reality of energy eigenvalues of edge states. A question arises. Are these edge states topological? In the Hermitian graphene, these edge states are robust against particle-hole symmetry preserving disorder. However, particle-hole symmetry is spontaneously broken for the edge states even if the bulk states are still particle-hole symmetric in the non-Hermitian case. To what kind of symmetry preserving disorder are the edge states in non-Hermitian graphene immune? Below, we answer this question.
\begin{figure}[t]
\includegraphics[width=4.25cm]{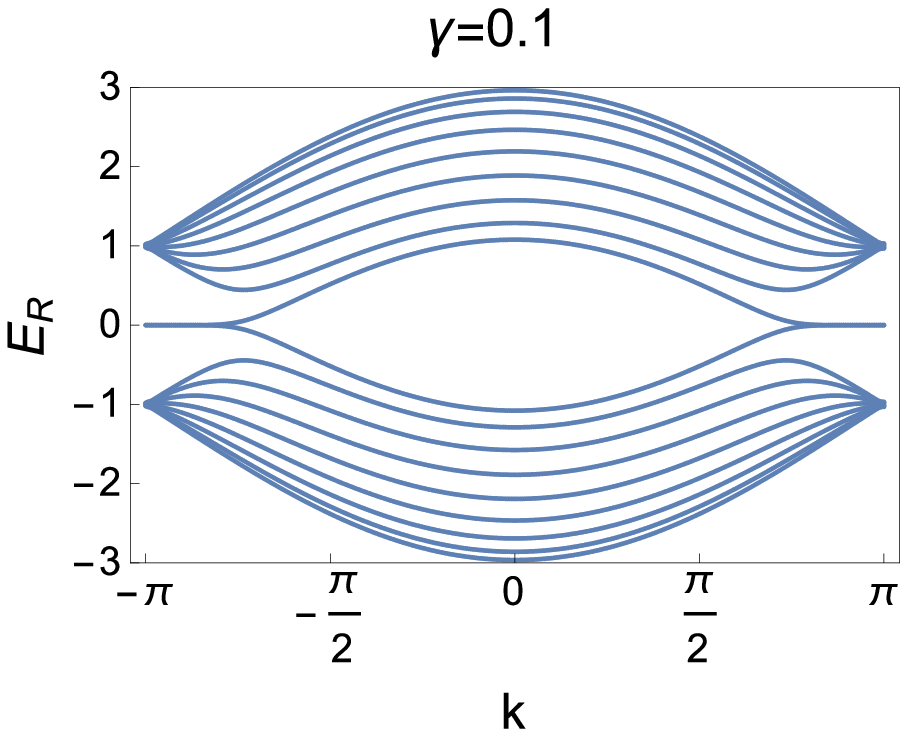}
\includegraphics[width=4.25cm]{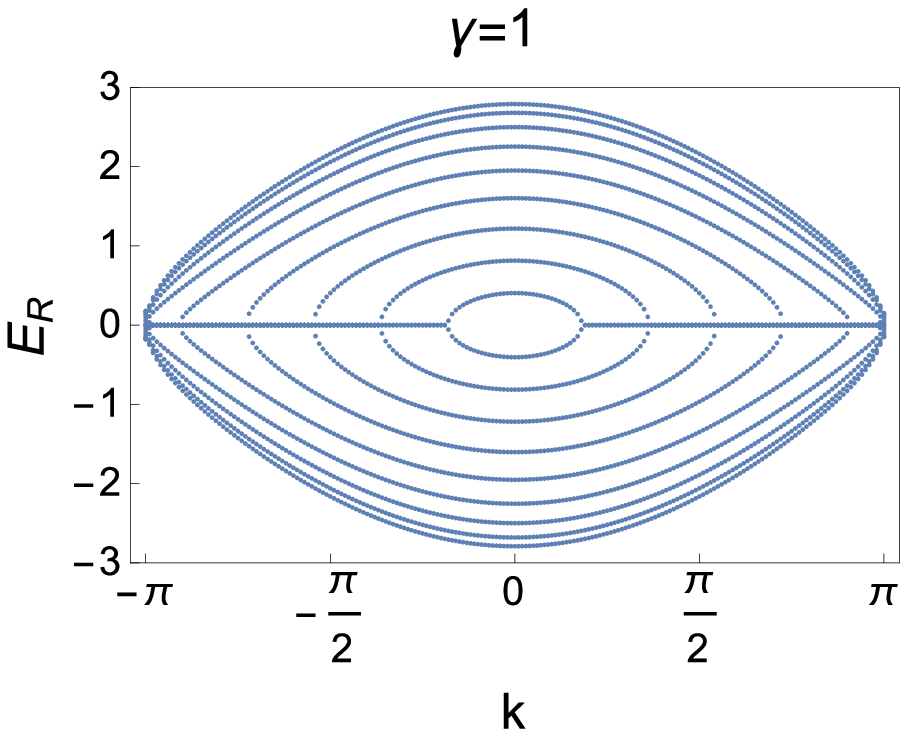}
\includegraphics[width=4.25cm]{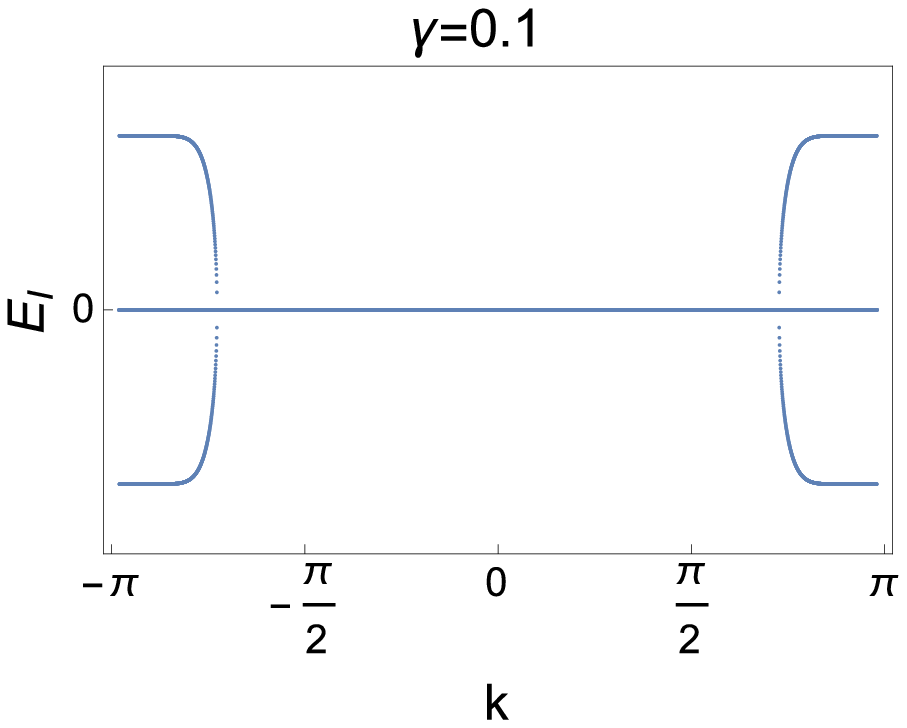}
\includegraphics[width=4.25cm]{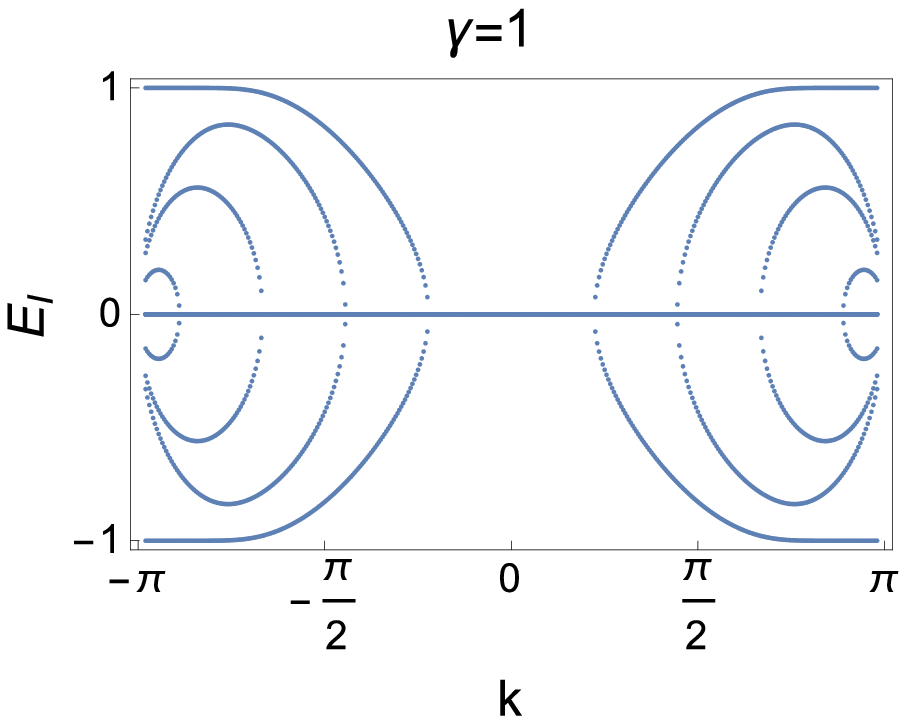}
\caption{ The real $\ds{E_R}$ (upper panel) and imaginary $\ds{E_I}$ (lower panel) parts of the energy eigenvalues for $\ds{\gamma=0.1}$ and $\ds{\gamma=1}$ as a function the wave vector $k$ along the zigzag edge. For a small value $\ds{\gamma=0.1}$, two edge states have purely imaginary energy eigenvalues while all the bulk states have real valued energy eigenvalues. For an intermediate value $\ds{\gamma=1}$,  a few number of bulk states have also complex energy eigenvalues in addition to the edge states.}
\end{figure}

\subsection{Spontaneously Broken Symmetry Protection}

As pointed above, exceptional rings whose centers are the Dirac points occur in the presence of gain and loss. The points on the exceptional rings corresponds to the states with zero energy eigenvalues. Inside these exceptional rings, energy eigenvalues are purely imaginary. The $\ds{\mathcal{PT}}$ and $\ds{\mathcal{C}}$ symmetries are spontaneously broken inside these exceptional rings while these symmetries are preserved outside of the exceptional rings. Therefore, the edge states for small values of $\gamma$ have no $\ds{\mathcal{PT}}$ and $\ds{\mathcal{C}}$ symmetries while all the bulk states have these symmetries.  In the theory of topological insulators, the edge modes are known to resist to symmetry-preserving perturbations. In our case, particle-hole symmetry is spontaneously broken for the edge states. So, which symmetry protects the edge states is unclear. Here, we propose an idea that these edge states with purely imaginary energy eigenvalues survive as long as the particle-hole symmetry is spontaneously broken in the system. In this case, the energy eigenvalues of these edge states don't change even in the presence of disorder (or pertubation) that does not restore spontaneously broken particle-hole symmetry. However,  the real and imaginary parts of the energy eigenvalues of the bulk states are sensitive to such disorder. Note that if the disorder breaks the particle-hole symmetry of the Hamiltonian (1), the idea of the spontaneously broken particle-hole symmetry is meaningless. Therefore, we are interested in disorder that respects particle-hole symmetry of the full Hamiltonian (1) and not be strong enough to restore the spontaneously broken particle-hole symmetry around the Dirac points. This {\it{spontaneously broken symmetry protection}} is unique to non-Hermitian systems. \\
\begin{figure}[t]
\includegraphics[width=3.9cm]{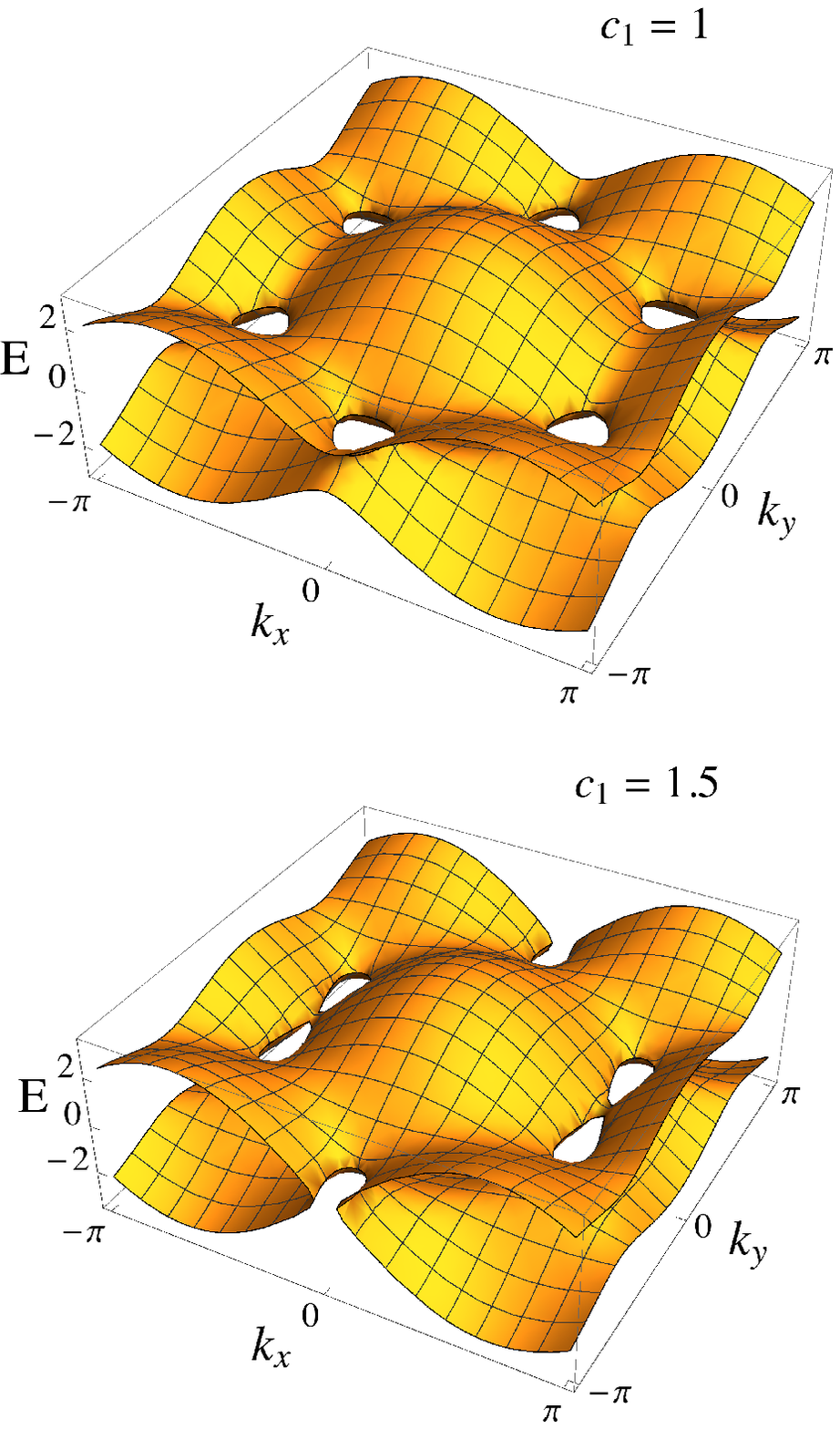}
\includegraphics[width=3.9cm]{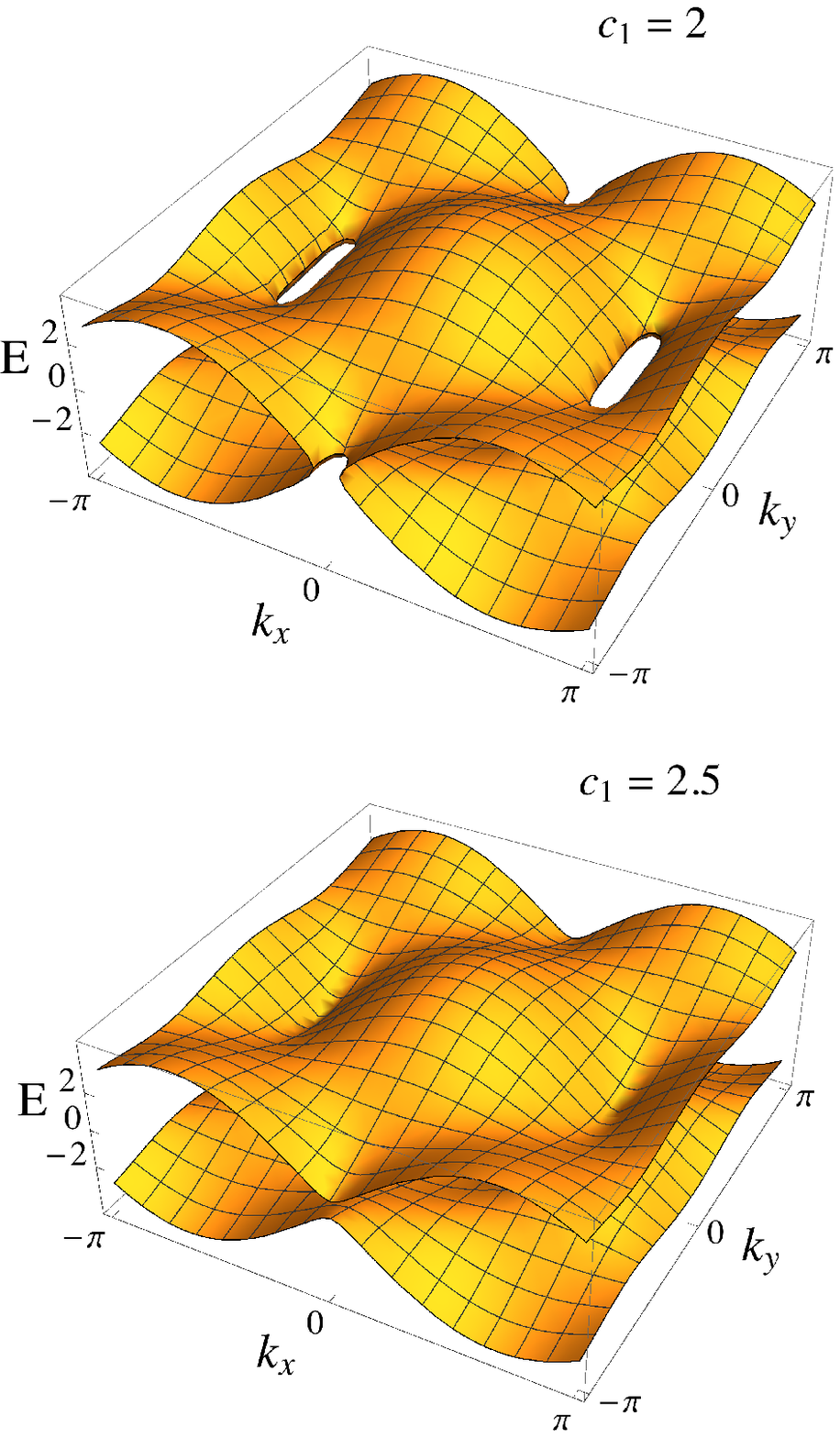}
\caption{ Band structure diagrams for compressed honeycomb photonic lattice described by the Hamiltonian (1) with (\ref{murygfD}). The region where the energy is complex valued is depicted by a hole in the figure. The borders of these holes form exceptional rings. As $c_1$ is increased for fixed $c_2$, exceptional rings move towards each other in pairs while shrinking. At a $\gamma$-dependent critical value, exceptional rings merge and destructs each other. In this case, the $\ds{\mathcal{PT}}$ and $\ds{\mathcal{C}}$ symmetries are restored again  }
\end{figure}
To illustrate our idea, we consider compressed graphene structure, which preserves the particle-hole symmetry. In this case, $d_x$ and $d_y$ in (\ref{mqasdfv2D}) become
\begin{eqnarray}\label{murygfD}
d_x+id_y=c_1~e^{-ik_x }+2c_2~e^{ i  \frac{3}{2} k_x}  ~ \cos(\frac{\sqrt{3}}{2}  k_y )  
\end{eqnarray}
where $c_1$ and $c_2$ are real valued free parameters describing compression. The regular graphene structure is obtained when $c_1=c_2=1$. Consider first the Hermitian case, $\gamma=0$. Since each of the Dirac points in the regular graphene has a Berry phase of $\mp\pi$, they are topologically protected against weak compressions. In fact, they move towards each other as the graphene is compressed. At a critical value $c_1=2c_2$, the Dirac points meet and the Berry phases $\mp\pi$ annihilate each other. The system becomes gapped and topologically trivial if $\ds{c_1>2c_2}$ \cite{grphnedge03}. In the presence of the gain and loss, $\gamma\neq0$, exceptional rings are generated from the Dirac points as we discussed above. These exceptional rings are still topologically protected. Assume that $c_1$ is varied for fixed $c_2=1$. Then the exceptional rings move towards each other in pair as can be seen from the Fig.3, where the exceptional rings are depicted as holes. We stress that the region where the $\ds{\mathcal{C}}$ symmetry is spontaneously broken move together with the exceptional rings. However, they shrink as they move. At some $\gamma$-dependent critical value, they merge and exceptional rings disappear. In other words, the $\ds{\mathcal{PT}}$ and $\ds{\mathcal{C}}$ symmetries are no longer spontaneously broken (the two symmetries are restored in the whole Brillouin zone) and the band gap is opened. In this case, edge states with purely imaginary energy along the zigzag edge disappear since the perturbation is strong enough to restore spontaneously broken particle-hole symmetry.\\
Let us briefly mention some other kinds of disorders. Consider a particle-hole symmetry preserving off-diagonal disorder, which can be introduced into our numerical computation by adding small random values to the tunneling between the blue and red sites in Fig-1. Since the exceptional rings are just deformed but not destroyed with this kind of disorder, the edge state survives and remain closely localized on the zig-zag edge. The real and imaginary parts of the energy eigenvalues for the edge states resist to disorder. i.e., they don't change even in the presence of the disorder although they change for the bulk states. Finally, consider now particle-hole symmetry breaking diagonal disorder, which can be introduced into our numerical computation by adding small random on-site potentials. Since the particle-hole symmetry is broken, it is meaningless to talk about spontaneously broken particle-hole symmetry. Therefore, even for very small values of such disorder, edge states disappear. \\
Another question arises. Are our findings specific to the graphene structure or can they be generalized to other particle-hole symmetric systems? Below, we discuss this issue briefly.

\section{Discussion}

In \cite{grphnedge01}, it was shown that zero-energy edge states exist for a class of particle-hole symmetric Hamiltonian. Consider a class of $1D$ or $2D$ Hermitian Hamiltonian $ \ds{\mathcal{H} (k)=d_x(\textbf{k})~\sigma_x+d_y(\textbf{k})~\sigma_y}$ that has particle-hole symmetry $\ds{\mathcal{C}=\sigma_z~  \mathcal{K}  }$. Assume that this Hermitian Hamiltonian supports zero energy edge states. Let us now add gain and loss to this Hamiltonian $\ds{\mathcal{H} (k)=d_x(\textbf{k})~\sigma_x+d_y(\textbf{k})~\sigma_y+i~{{\gamma}}~ \sigma_z}$, where $\gamma$ is the non-Hermitian degree. There are two important systems. i-) $\it{Gapless ~ systems:}$ In this case, particle-hole symmetry is always spontaneously broken at the points where energy gap of the Hermitian part is zero and exceptional points  (or rings) may occur unless the non-Hermitian degree doesn't exceed a critical number. As discussed above for the specific graphene problem, one can find edge states with purely imaginary energy eigenvalues in the corresponding finite lattice with open edges. ii-) $\it{Gapped ~ systems:}$  In this case, particle-hole symmetry is spontaneously broken for some particular values of $\textbf{k}$ only when $\gamma$ exceeds a critical value, $\gamma_c$. As an example, consider the following complex extension of the Su-Schrieffer-Heeger (SSH) model $\mathcal{H} (k)=\left(\nu+\omega\cos(k) \right)\sigma_x+\omega\sin(k)\sigma_y+i\gamma \sigma_z$, where the crystal momentum $k$ runs over the first Brillouin zone, $-\pi<k<\pi$ and the real-valued positive parameters $\ds{\nu>0}$, $\ds{\omega>0}$ are hopping amplitudes. This Hamiltonian has particle-hole symmetry $\ds{\mathcal{C}=\sigma_z~  \mathcal{K}  }$. We emphasize that particle-hole symmetry is preserved unless $\gamma<\gamma_c=|\omega-\nu|$ for periodical system. In the case of finite SSH chain with open edges, the picture changes considerably. The air is topologically trivial and the finite SSH chain must be topologically nontrivial to observe topological edge states. In other words, topological phase transition occurs at the edges ( the band gap should close and reopen for the topological phase transition). The band gap closing implies that particle-hole symmetry is spontaneously broken at the edges for any value of $\gamma$. Therefore, edge states with purely imaginary eigenvalues appear in the system. These edge states are protected by spontaneously broken particle-hole symmetry as discussed above. Note that some authors found other ways to circumvent this problem to get stable (real-valued energy) topological edge states \cite{cem0001,sondeney1}.

\section{Conslusion}

We have studied edge states in photonic graphene with balanced gain and loss. We have shown that edge states with purely imaginary eigenvalues appear in the system. The particle-hole symmetry is spontaneously broken for the edge states. However, the edge states are still protected. We claim that the protection is the result of spontaneously broken particle-hole symmetry. In this case, the energy eigenvalues of the edge states resist to disorder. The edge states are topological in the sense that the exceptional rings is robust with respect to perturbations. If the disorder is too strong to restore the spontaneously broken particle-hole symmetry, then such protection of the edge states is lost. We stress that particle-hole symmetry leads to zero energy edge states while spontaneously broken particle-hole symmetry leads to edge states with purely imaginary energy. This spontaneously broken symmetry protection is unique to non-Hermitian systems. Our analysis is generic and can be applied to gapped and gapless particle-hole symmetric Hamiltonians. It is worth generalizing our analysis to systems with spontaneously broken time-reversal and chiral symmetric non-Hermitian systems.

\end{document}